\documentclass[11pt]{article}
\usepackage{amsfonts}
\usepackage{amsmath,amssymb}
\usepackage{pstricks}
\usepackage{pst-node}
\usepackage{pst-tree}
\usepackage{pst-all}

\baselineskip \advance \textheight by \topskip \textwidth 470pt
\makeatletter \@addtoreset{equation}{section}

\makeatother
\def\Z{\mathbb Z}

\def\R{\mathbb R}

\begin{document}

\title{Peculiarities of the hidden nonlinear
supersymmetry  of P\"{o}schl-Teller system in the light of Lam\'e
equation}
\author{\textsf{Francisco Correa}\thanks{%
fco.correa.s@gmail.com} \textsf{$\,$ and  Mikhail S. Plyushchay}%
\thanks{%
mplyushc@lauca.usach.cl} \\
%EndAName
{\small \textit{Departamento de F\'{\i}sica, Universidad de
Santiago de Chile, Casilla 307, Santiago 2, Chile}} }
\date{}
\maketitle

\begin{abstract}
A hidden nonlinear bosonized supersymmetry was revealed recently in
P\"{o}schl-Teller and finite-gap Lam\'{e} systems. In spite of the
intimate relationship between the two quantum models, the hidden
supersymmetry in them displays essential differences. In particular,
the kernel  of the supercharges of the P\"oschl-Teller system,
unlike the case of Lam\'e equation, includes nonphysical states. By
means of Lam\'e equation, we clarify the nature of these peculiar
states, and show that they encode essential information not only on
the original hyperbolic P\"oschl-Teller system, but also on its
singular hyperbolic and trigonometric modifications, and reflect the
intimate relation of the  model to a free particle system.
\end{abstract}

\section{\protect\bigskip Introduction}

Both on the classical and quantum levels, symmetries are behind
the special properties of the systems. Sometimes symmetries appear
in a hidden form like it happens, for instance, in the case of a
spontaneously broken symmetry. Another, well known mechanical
example is provided by the model of hydrogen atom being the
quantum analog of the Kepler problem, in which a hidden symmetry
associated with Laplace-Runge-Lenz vector underlies a specific
degeneration of the spectrum \cite{runge}. Recently, it was found
\cite{susypt, lame} that some well  studied quantum mechanical
systems exhibit a bosonized supersymmetry  \cite{boso} in a hidden
form. The hidden supersymmetry manifests explicitly the main
characteristics of the systems, and may have a linear, or
nonlinear \cite{para} character. Hidden supersymmetry of linear
form appears in the bound state Aharonov-Bohm effect and the Dirac
delta potential problems \cite{susypt}. The  pure bosonic quantum
P\"{o}schl-Teller (\textit{PT}) \cite{susypt} and Lam\'{e}
\cite{lame} systems display hidden supersymmetry of nonlinear
form.

The \textit{PT} and Dirac delta quantum problems are special
limits of the Lam\'e equation. In the form of the periodic quantum
problem, the latter system underlies  diverse models and
mechanisms in field theory \cite{field,LMT}, nonlinear wave
physics \cite{Solitons}, cosmology \cite{PreHeat}, condensed
matter physics  \cite{Suth, AGI,GIN} and statistical mechanics
\cite{bazh}. Jacobian form of Lam\'e equation \cite{ww,arscott}
usually used in physics is
\begin{equation}
H_{j}^{L}\Psi
=0,~~H_{j}^{L}=-\frac{d^{2}}{dx^{2}}+j(j+1)k^{2}\,\mathrm{sn}
^{2}(x,k)+c,  \label{lame}
\end{equation}
where $\mathrm{sn}(x,k)\equiv \,\mathrm{sn}\,x$  is the Jacobi
elliptic sine function, $k$, $0<k<1$, is the modular elliptic
parameter,  while $j$ and  $c=c(j,k)$ are  real constants. Eq.
(\ref{lame}) can be treated as a Schr\"{o}dinger one-dimensional
equation with a doubly periodic potential, in which $-c$ has a
sense of an eigenenergy.

When the modular parameter takes its limiting values, we obtain two
different
 systems. For $k=0$ (and finite $j$), potential term disappears from
$H_{j}^{L}$, and the Hamiltonian corresponds to a free particle.
Meanwhile in the limit $k=1$, the real period of the potential
turns into infinity, and (\ref{lame}) is reduced to the
P\"{o}schl-Teller system, $ H_{j}^{L}\rightarrow
H_{j}^{\textit{PT}}=-\frac{d^{2}}{dx^{2}}-j(j+1)\mathrm{sech}
^{2}x+c^{\prime }$. Since the periods of elliptic function
$\mathrm{sn}^2\, x$ are $2\mathbf{K}$ and $2i\mathbf{K}'$, while
$\mathrm{sech}^2\, x$ has the imaginary period $i\pi$, the
potential of Lam\'e system can be treated as a certain periodic
superposition of the \textit{PT} potentials \cite{Dunnes},
\begin{equation}
H_{j}^{L}=-\frac{d^{2}}{dx^{2}}-j(j+1)\left( \frac{\pi
}{2\mathbf{K}^{\prime }}\right)
^{2}\sum_{l=-\infty }^{\infty }\mathrm{sech}^{2}\left( \frac{\pi }{%
2\mathbf{K}^{\prime }}\left[ x-2l\mathbf{K}\right] \right)
+j(j+1)\frac{\mathbf{E'}}{\mathbf{K'}}+c.
\end{equation}

What makes the Lam\'e and P\"oschl-Teller models to be
particularly interesting are those remarkable properties appearing
when the parameter $j$ takes integer values $n$ (for the sake of
definiteness we assume $n=1,2,\ldots $). For these special values
Lam\'e equation describes  a \emph{finite-gap} quantum periodic
system, while \textit{PT} system is \emph{reflectionless}.

Behind these special properties, there emerges the  hidden
nonlinear supersymmetry. For $j=n$, both systems have nontrivial
integrals of motion in the form of differential operators of order
$2n+1$. These corresponding odd integrals of motion, $Q_{n}^{L}$
and $Q_{n}^{\textit{PT}}$, together with reflection (being an
obvious symmetry) play the role of the supercharge and grading
operators of the hidden nonlinear supersymmetry \cite{susypt,
lame}.

Though the nonlinear supersymmetry in both models has a somewhat
similar structure, there are essential differences between its
realizations. In both systems all the physical singlet states are
annihilated by the supercharges, but in the \textit{PT}, unlike
the Lam\'e case, the supercharge has also non-normalizable,
nonphysical (formal) zero modes. This difference is reflected in
the nonlinear superalgebraic structure. The square of the
supercharge in both systems gives a polynomial of order $2n+1$ in
corresponding Hamiltonian operator. For (\ref{lame}), we get the
spectral polynomial with all the roots to be simple and equal to
the energies of the edges of the allowed bands. However, for the
\textit{PT} system, the polynomial has $n$ double roots associated
with the bound states, while one simple root corresponds to the
lowest, singlet state of the continuous spectrum. Having in mind
these nonlinear superalgebraic relations between corresponding
supercharges and Hamiltonians,  in Lam\'e (L) and \textit{PT}
systems with $j=n$ we deal, respectively, with non-degenerate and
degenerate hyperelliptic curves of genus $g=n$,
\begin{equation}\label{hyper}
    L:\,\, y^2=\prod_{k=0}^{2n+1}(z-z_k),\qquad
    \textit{PT}:\,\, y^2=(z-z_{2n+1})\prod_{k=0}^{n}(z-z_k)^2,\quad
    z_k\neq z_{k'}.
\end{equation}

The existence of these differences gives rise to the following
questions: what is the relation between the two systems from the
point of view of the hidden supersymmetry, and, on the other hand,
if the supercharges are fundamentals objects that contain all the
information on the systems, what is the origin and nature of
nonphysical states from the \textit{PT} supercharge kernel? It is
the purpose of the present article to answer these questions.

The paper is organized as follows. In Section 2 the hidden
nonlinear supersymmetric structure of the \textit{PT} and Lam\'e
systems is reviewed, and the structure of their supercharge
kernels and relation between them is discussed. In  Section 3 the
nature of non-normalizable, nonphysical states of the \textit{PT}
supercharge kernel is clarified, and their origin in the light of
Lam\'e equation is investigated. We conclude in Section 4 with a
brief summary.

\section{\protect\bigskip Hidden supersymmetry of Lam\'e and \textit{PT} systems}

A part of the spectrum of the Lam\'e and \textit{PT} systems with
$j=n$ is doubly degenerated. For (\ref{lame}) this corresponds to
the energies of the quasiperiodic (Bloch-Floquet) states of the
internal part of the valence  and conduction bands, while in the
\textit{PT} system these are the energies of the scattering states
(except the lowest one). Double degeneration of the energy levels
is a characteristic feature of the $N=2$ supersymmetry generated
by two supercharges. On the other hand, there are $2n+1$ singlet
states corresponding to the edges of the allowed bands in Lam\'e
system, and $n+1$ singlets corresponding to $n$ bound states plus
one nondegenerate lowest state of the scattering sector in
\textit{PT} system. In both cases the number of singlet states is
greater than one, that is typical for nonlinear supersymmetry
\cite{para, nonli}.

The nonlinear supersymmetry in both systems is generated by the
local, $Q_{n}^{L,\textit{PT}}=Q_{n}$,  and nonlocal (due to a
nonlocal nature of $R$), $\tilde{Q} _{n}=iRQ_{n}$, supercharges,
\begin{equation}
\lbrack Q_{n},H_{n}]=[\tilde{Q}_{n},H_{n}]=0,\quad \{Q_{n},\tilde{Q}_{n}\}=0,
\label{brac}
\end{equation}%
\begin{equation}
Q_{n}^{2}=\tilde{Q}_{n}^{2}=P_{2n+1}(H_{n}),  \label{poly}
\end{equation}%
where $P_{2n+1}(H_{n})$ is a polynomial of order  $2n+1$ in
Hamiltonian $H_{n}=H_{n}^{L,\textit{PT}}$ and  $R$ is the
reflection operator $R\Psi (x)=\Psi (-x)$ identified as the
grading operator, $[R,H_{n}]=0,$
$\{R,Q_{n}\}=\{R,\tilde{Q}_{n}\}=0$, $R^{2}=1$. {}From the
structure of the nonlocal supercharge $\tilde{Q}_{n}$ it is clear
that its kernel coincides with that of the local supercharge
$Q_n$.

The $2n+1$ nondegenerate states of the edges of the allowed bands
of the system (\ref{lame}) are given by the Lam\'e polynomials
\cite{ww}, \cite{arscott} of the form
\begin{equation}
\mathrm{sn}^{r}\,x\,\mathrm{cn}^{s}\,x\,\mathrm{dn}^{t}\,x\,F_{p}(\mathrm{sn}^{2}\,x),
\label{lampoly}
\end{equation}
where $F_{p}(\mathrm{sn}^{2}\,x)$ is a polynomial of order $p$ in
$\mathrm{sn} ^{2}\,x,$ $r,s,t=0$ or $1$, and $r+s+t+2p=n$. The
states  (\ref{lampoly}) are annihilated by $Q_{n}^{L}$, and thus
form the supercharge kernel $K_{n}^{L}=\ker Q_{n}^{L}$.

For $n=0$ the Lam\'e system reduces to a trivial case of  a free
particle, and the momentum operator $Q_{0}^{L}=-iD$,
$D=\frac{d}{dx}$, is identified as a supercharge. The first
nontrivial case corresponding to $n=1$ is characterized by the
supercharge
\begin{equation}
iQ_{1}^{L}=D^{3}+fD+\frac{1}{2}f^{\prime }, \label{P1}
\end{equation}
where $f^{\prime }=\frac{d}{dx}f,$ and $f$ is a doubly periodic
function being, up to a shift of the argument, the Weiesstrass
elliptic $\wp$-function with periods $\omega_1=2\textbf{K}$ and
$\omega_2=2i\textbf{K}^{\prime }$,
\begin{equation}
f:=1+k^{2}-3k^{2}\,\mathrm{sn}^{2}x=-3\wp (x+i\mathbf{K}^{\prime
}).  \label{efe}
\end{equation}
The supercharge for arbitrary even (odd) $n$ is constructed
recursively via $Q^L_0$ ($Q^L_1$),
\begin{equation}
Q^L_{n}=\Lambda _{n}Q^L_{n-2},\quad n>1,  \label{LamQ}
\end{equation}
where  $\Lambda _{n}$ is a differential operator of order $4$,
\begin{eqnarray}
&\Lambda _{n} =D^{4}+\left[ ~2n(n-1)+1\right] fD^{2}+\left[ \frac{4}{3}%
\left( n-1\right) \left( n-\frac{1}{2}\right) \left( n+\frac{3}{2}\right) +%
\frac{1}{2}\right] f^{\prime }D+&\nonumber \\
&(n-1)^2\left[ \frac{2}{3}\left(\left(
n+\frac{1}{2}\right)^{2}+\frac{1}{2}\right) f^{\prime \prime }+
n^2f^{2}\right].& \label{Lamj}
\end{eqnarray}

Though Lam\'e polynomials (\ref{lampoly}) as well as the
corresponding energies $E_{n,l}^L$ of the band edges can be found in
analytic form for $n\leq 8$ \cite{maier}, in correspondence with
recurrent structure of the supercharges, their kernels can be
presented explicitly in general case in terms of monomials in
$\mathrm{sn}\,x$, $\mathrm{cn}\,x$ and $\mathrm{dn} \,x$
\cite{lame},
\begin{equation}
K_{n}^{L}=\left\{
K_{n-2}^{L},~\mathrm{dn}^{n}\,x,~\mathrm{cn}\,x\,\mathrm{dn}
^{n-1}\,x,~\mathrm{sn}\,x\,\mathrm{dn}^{n-1}\,x,~\text{\textrm{cn}}\,x\,
\mathrm{sn}\,x\, \mathrm{dn}^{n-2}\,x\right\}, \quad n>1,
\label{lameker}
\end{equation}
where $K_{0}^{L}=1$, $K_{1}^{L}=\{\mathrm{dn}\,x,\,\mathrm{cn}\,x,\,
\mathrm{sn}\,x\}$. The monomial elliptic functions of the kernel are
certain linear combinations of Lam\'e polynomials.

The square of the supercharge (\ref{poly}) is given by the Lam\'e
spectral polynomial,
\begin{equation}
P_{2n+1}^{L}(H^{L}_n)=\prod_{l=0}^{2n}(H^{L}_n-E_{n,l}^{L}),\label{Lapoly}
\end{equation}
where $E_{n,l}^{L}$, $l=0,\ldots ,2n$,  are the eigenvalues of the
edges of the allowed bands.

The limit $k=1$ preserves the hidden supersymmetry, and transforms
the periodic finite-gap Lam\'e equation (\ref{lame}) into
reflectionless P\"oschl-Teller system,
\begin{equation}
H_{n}^{L}\underset{k=1}{\longrightarrow
}H_{n}^{\textit{PT}}=-D^{2}-n(n+1)\,\mathrm{
sech}^{2}x+n^{2}=\mathcal{D} _{n}^{\dagger }\mathcal{D}_{n},
\label{pt}
\end{equation}
where $\mathcal{D}_{n}=-\mathcal{D}_{-n}^{\dagger
}=\frac{d}{dx}+n\tanh x$. The states
\begin{equation}
\psi _{n,l}={\cal D}^l_{-n}\cosh ^{l-n}x, \quad  l=0,1,\ldots,n,
\label{PTbs}
\end{equation}
with
\begin{equation}\label{DDl}
    {\cal D}^0_{-n}=1,\quad
     {\cal D}^1_{-n}={\cal D}_{-n},\quad
{\cal D}^l_{-n}=\mathcal{D}_{-n}\mathcal{D}_{-n+1}\ldots \mathcal{D}
_{-n+l-1},\quad l=2,\ldots ,n,
\end{equation}
represent $n$ bound states corresponding to $l=0,...,n-1$, while the
state with $l=n$ is the lowest state from the continuous part of the
spectrum. Their energies are given by
\begin{equation}
E^{\textit{PT}}_{n,l}=n^{2}-(n-l)^{2},\qquad l=0,\ldots ,n.
\label{EnPT}
\end{equation}
A specific choice of the constant shift in (\ref{pt}) corresponds
to zero energy value of the ground state $\psi_{n,0}$. These $n+1$
singlet eigenstates of the Hamiltonian constitute a part of the
kernel $K_{n}^{\textit{PT}}=\ker Q_{n}^{\textit{PT}}$ of the
supercharge. The latter can be obtained directly from (\ref{LamQ})
by taking the limit $k=1$. This supercharge can be represented in
more elegant form \cite{susypt},
\begin{equation}\label{QPTel}
Q^{\textit{PT}}_n=i\mathcal{D}_{-n}\mathcal{D}_{-n+1}...\mathcal{D}_{n}.
\end{equation}
Its square gives a  corresponding polynomial (\ref{poly}),
\begin{equation}
P_{2n+1}^{\textit{PT}}(H_{n}^{\textit{PT}})=(H_{n}^{\textit{PT}}-E_{n,n}^{\textit{PT}})\prod_{l=0}^{n-1}\left(
H_{n}^{\textit{PT}}-E_{n,l}^{\textit{PT}}\right)
^{2}.\label{polyPT}
\end{equation}

This polynomial is a $k=1$ limit of the Lam\'e spectral polynomial
(\ref{Lapoly}), but unlike the latter, it has $n$ double roots
corresponding to the singlet bound states energies, while its one
simple root corresponds to the energy of the singlet lowest state
from the continuous spectrum. All the associated $n+1$ singlet
eigenstates (\ref{PTbs}) are annihilated by the supercharge.
However, supercharge (\ref{QPTel}) is the differential operator of
the order $2n+1$, and its complete kernel (without taking into
account the question of normalizability of the states) of dimension
$2n+1$ is given recursively as
\begin{equation}
K_{n}^{\textit{PT}}=\left\{
K_{n-2}^{\textit{PT}},~\mathrm{cosh}^{-n}\,x,~\mathrm{cosh}^{-n}\,x\sinh
x,~\cosh ^{n-2}\,x\sinh x,~\cosh ^{n}x\,\right\},\quad n>1,
\label{ptker}
\end{equation}
where
$K_{0}^{\textit{PT}}=1,K_{1}^{\textit{PT}}=\{\mathrm{sech}\,x,\,\tanh\,
x,\,\cosh\, x\}$. It is spanned by the states
\begin{equation}\label{KPTn}
    K_n^{\textit{PT}}=\{\cosh^s x\sinh^r x,\quad
    s=-n,-n+2,\ldots, n-2r,\quad r=0,1\},
\end{equation}
and  can be rearranged as follows,
\begin{equation}
K_{n}^{\textit{PT}}=\left\{ \mathcal{\hat{K}}_{n}^{\textit{PT}},~\mathcal{\kappa }_{n}^{\textit{PT}},~%
\mathcal{\tilde{K}}_{n}^{\textit{PT}}~\right\} ,\quad \dim
\mathcal{\hat{K}}_{n}^{\textit{PT}}=\dim
\mathcal{\tilde{K}}_{n}^{\textit{PT}}=n,\quad \dim \mathcal{\kappa
}_{n}^{\textit{PT}}=1.
\end{equation}
Here $\mathcal{\hat{K}}_{n}^{\textit{PT}}$ corresponds to the
normalizable (with respect to the ordinary scalar product  on
$\R^1$) functions,

\begin{equation}\label{KPThat}
    \mathcal{\hat{K}}_{n}^{\textit{PT}}=\{\cosh^s x\sinh^r x,\quad
    r=0,1,\quad\left\{
\begin{array}{l}
s=-n,-n+2,\ldots,-2,\quad n=2m>0\\
s=-n,-n+2,\ldots,-(2r+1),\quad n=2m+1
\end{array}
\right.\},
\end{equation}
being linear combinations of the $n$ bound states, while
\begin{equation}\label{kappaPT}
\mathcal{\kappa}_{n}^{\textit{PT}}= \left\{
\begin{array}{l}
1,\quad   n=2m, \\
\tanh x,\quad   n=2m+1,
\end{array}
\right.
\end{equation}
are linear combinations of the singlet scattering state
$\psi_{n,n}$ and bound states $\psi_{n,l}$,  $l=0,\ldots,n-1$.
These are the $n+1$ states corresponding to the $k=1$ limit of the
supercharge kernel (\ref{lameker}) of Lam\'e system,
$K_{n}^{L}\underset{k=1}{\longrightarrow }\left\{
\mathcal{\hat{K}}_{n}^{\textit{PT}}, \mathcal{\kappa
}_{n}^{\textit{PT}}\right\}.$ In this limit, the period of Lam\'e
equation tends to infinity, the valence bands shrink, and two edge
states (and their energies) of the same band converge smoothly in
one bound state (and corresponding energy) of P\"oschl-Teller
system. The states of the continuous band of (\ref{lame}) in this
limit are transformed into the states of continuous spectrum of
(\ref{pt}), and the singlet edge state of the conduction band is
transformed into the first (lowest) singlet state of the
continuous spectrum. {}From another point of view, since in the
limit $k=1$ Jacobi functions
 $\mathrm{cn}\,x$ and $\mathrm{dn}\,x$
are reduced to the same function $\mathrm{sech}\,x$, two different
Lam\'e polynomials are transformed into the same function in terms
of associated Legendre functions of the variable $\tanh\, x$
\cite{arscott}.

Therefore, from the point of view of the $k=1$ limit of Lam\'e
system, the origin of the non-normalizable  non-physical states
$\cosh^{s}x$, $s\geq 1$, and $\cosh ^{s'}x\sinh x$, $s'\geq 0$,
from the $n$-dimensional subspace
$\mathcal{\tilde{K}}_{n}^{\textit{PT}}$ of the total kernel
$K_{n}^{\textit{PT}}$ of the P\"oschl-Teller supersupercharges
seems to be mysterious.

\section{The nature and origin of $\mathcal{\tilde{K}}_{n}^{\textit{PT}}$}

Before investigating the question on the origin of the
non-normalizable nonphysical states of the supercharge kernel  from
the point of view of the associated Lam\'e system, we clarify their
nature within the framework of the P\"oschl-Teller system itself.

First we note that the  states of the complete kernel (\ref{KPTn})
with the same parity but different values of the parameter $s$ can
be related by the Hamiltonian operator,
\begin{equation}\label{kerker}
\left( H_{n}-E_{n,n-\left\vert s+r\right\vert }\right) \cosh^s
x\sinh ^r x=C_{s,n} \cosh ^{s-2}x\sinh^r x,
\end{equation}
where $s=-n,-n+2,...,n- 2r,$ $C_{s,n}=s(s-1)-n(n+1)$
 and
$E_{n,l}=E^{\textit{PT}}_{n,l}$ are the energies given by Eq.
(\ref{EnPT}). On the other hand, all the physical singlet states
(\ref{PTbs}) (both from the bound and continuous parts of the
spectrum) can  be produced by the action of the polynomial in
Hamiltonian operator on the states of the supercharge kernel
belonging to ${\hat{K}}_{n}^{\textit{PT}}$ or
$\kappa_n^{\textit{PT}}$,
\begin{equation}\label{psiker}
\prod\limits_{s=0}^{_{\frac{1}{2}(n-l)-1}}(H_{n}-E_{n,2s+r})\,
\cosh^{-l}x\sinh ^{r}x=\psi _{n,n-l+r}(x),
\end{equation}
where $l=n-2,n-4,\ldots, 2r$ for even $n$, and $l=n-2,n-4,\ldots,1$
for odd $n$, with $r=0,1$.

Combining relations (\ref{kerker}) and (\ref{psiker}) we conclude
that any physical singlet  state can be obtained from any
non-physical (exponentially increasing) state (\ref{KPTn}) of the
supercharge kernel by applying to it a certain polynomial operator
in the Hamiltonian.

The states of the complete supercharge kernel (\ref{KPTn}) have
also the following property. Let us identify their logarithmic
derivatives,
\begin{eqnarray}
&W_{0,s}=-\frac{d}{dx}\ln (\cosh^s x) =-s \tanh x,\quad
W_{1,s}=-\frac{d}{dx}\ln (\cosh^s\sinh x) =W_{0,s}-\coth x,&
\label{W01}
\end{eqnarray}
as superpotentials in the sense  of a usual linear supersymmetry,
and construct  the corresponding superpartner Hamiltonians
$H^\pm_{r,s}=-D^2+W^2_{r,s}\pm W'_{r,s}\,$, where we assume that
$s\in \Z $. Then we get
\begin{equation}
H_{0,s}^{+}=-D^{2}-s (s +1)\,\mathrm{sech}^{2}x+s ^{2},\qquad
H^-_{0,s}=H^+_{0,s-1}+2s-1\, , \label{H0s+}
\end{equation}
\begin{equation}\label{H1s-}
H_{1,s}^{+}=-D^{2}-s(s+1)\mathrm{sech}^{2}x+(s+1)^{2}+2\mathrm{cosech^{2}}x,
\qquad H^-_{1,s}=H^+_{0,s-1}+4s.
\end{equation}
Since $H^+_{0,s}=H^-_{0,-s}$, from the point of view of such a
construction both physical and nonphysical states of the complete
kernel of the supercharge are, in fact, equivalent, and every time
we produce either a  (shifted) reflectionless P\"oschl-Teller
system with the corresponding value of the constant parameter, or
a free particle ($H^+_{0,0}$, $H^-_{0,1}$, $H^-_{1,1}$), or the
generalized P\"oschl-Teller system ($H^+_{1,s}$) \cite{cooper}. In
particular, note that the states $\cosh^{\pm n} x$ generate
exactly the ``parent" system given by
$H^+_{0,n}=H^-_{0,-n}=H^{\textit{PT}}_n$. {}From relations
(\ref{H0s+}) it also follows the well known fact lying behind the
reflectionless property: the supersymmetric partner of the
\textit{PT} system with $n=1$ is a free particle, while the
\textit{PT} system with $j=n>1$ is related to a free particle via
usual, linear supersymmetry in $n$ steps \cite{cooper}.

Taking into account the relations $H^+_{0,s-1}\cosh^s
x=(1-2s)\cosh^s x$, $H^+_{0,s-1}(\cosh^s x\sinh x)=-4\cosh^s\sinh
x$, and $H^+_{0,-s}=H^+_{0,s-1}+2s-1$, we get
\begin{eqnarray}\label{HPTchn}
    &H^{\textit{PT}}_n\cosh^{-n} x=0,\qquad
    H^{\textit{PT}}_n(\cosh^{-n}\sinh x)=(2n-1)\cosh^{-n}x\sinh x,&\\
    &H^{\textit{PT}}_{n-1}\cosh^nx=-(2n-1)\cosh^n
    x,\qquad
    H^{\textit{PT}}_{n-1}(\cosh^n x\sinh x)=-4n\cosh^n x\sinh
    x.&\label{HPTchsh}
\end{eqnarray}
While two physical states from (\ref{KPTn}) with $s=-n$, $r=0,1$
are the first (ground)  and the second singlet eigenstates of the
system $H^{\textit{PT}}_n$, other physical states of the kernel
(\ref{KPTn}) are the first and the second singlet eigenstates of
the \textit{PT} systems $H^{\textit{PT}}_s$ with corresponding
values of the parameter $0<s<n$. Moreover, according to Eq.
(\ref{HPTchsh}), the \emph{nonphysical} states of the kernel
$K^{\textit{PT}}_n$ can also be identified as
\emph{non-normalizable} eigenstates of the P\"oschl-Teller (or
shifted free particle) systems with corresponding \emph{negative}
eigenvalues.

Let us apply a Wick rotation to the \textit{PT} potential to see
further evidence for importance of the nonphysical states of the
\textit{PT} supercharge kernel. The rotation  can be realized by
``restoring" a frequency parameter $\omega$ in the potential term,
$H_{n}^{\textit{PT}}=-D^{2}-n(n+1)\,\omega
^{2}\mathrm{sech}^{2}\omega x+n^{2}\omega ^{2}$, subsequent
substitution $\omega\rightarrow i\omega$, and then putting again
$\omega=1$. In this way we get the trigonometric
 P\"oschl-Teller system,
$H^{\textit{PT}}_n\rightarrow
\tilde{H}^{\textit{PT}}_n=-D^{2}+n(n+1)\mathrm{sec} ^{2} x-n^{2}$.
Under such a procedure, nonphysical states $\cosh^n x$ and
$\cosh^n x\sinh x$ are transformed into the trigonometric
counterparts $\cos^n x$ and $\cos^n x\sin x$ belonging to the
kernel of the transformed supercharge operator. But now they are
physical states of the trigonometric \textit{PT} system with
$j=n-1$, which, in correspondence with Eq. (\ref{HPTchsh}), are
the two first (lower) eigenstates of the
$\tilde{H}^{\textit{PT}}_{n-1}$. Analogously, other nonphysical
states of the supercharge kernel of the hyperbolic \textit{PT}
system are transformed into physical states of the supercharge
kernel of the trigonometric \textit{PT} system. On the other hand,
normalizable states from physical part of the supercharge kernel
are transformed into nonphysical states (which violate necessary
boundary conditions at $x=\pm \pi/2$) of the trigonometric
\textit{PT} supercharge kernel. \vskip 0.2cm

Let us return  to the  Lam\'e system. As we have seen, the limit
$k=1$ produces from the Lam\'e supercharge kernel only the
physical states of the P\"oschl-Teller supercharge kernel. We
shall show that nonphysical states of the \textit{PT} supercharge
kernel can also be obtained in the same limit proceeding from the
Lam\'e system. For this we note that due to periodicity, a
constant real shift of the argument in the Lam\'e Hamiltonian
results just in a shift of the potential along a real line, but
does not change the spectrum of the system and its special
properties. With this observation, let us shift the argument for a
half of the real period of the potential \cite{dunsuk},
$x\rightarrow x+\mathbf{K}$,
\begin{equation}
H_{n}^{L}\underset{x=x+\textbf{K}}{\longrightarrow
}{H}_{n}^{L+\mathbf{K}}=-D^{2}+n(n+1)\left[ 1-k^{\prime
}{}^{2}\,\mathrm{dn}^{-2}(x,k)\right] +c.  \label{isos}
\end{equation}
Shifting  the argument in the corresponding formulas for the
Lam\'e system, we get the supercharge $Q^{L+\mathbf{K}}_n$ for the
system (\ref{isos}) and its corresponding kernel
$K_{n}^{L+\mathbf{K}}$. The difference of the shifted system
(\ref{isos}) in comparison with the original one (\ref{lame}) is
that in both limits $k=0$ and $k=1$ it transforms into the free
particle but with different additive constants. The relationship
between \textit{PT}, Lam\'e and free particle systems is
summarized in Fig. 1.
\begin{figure}[h]
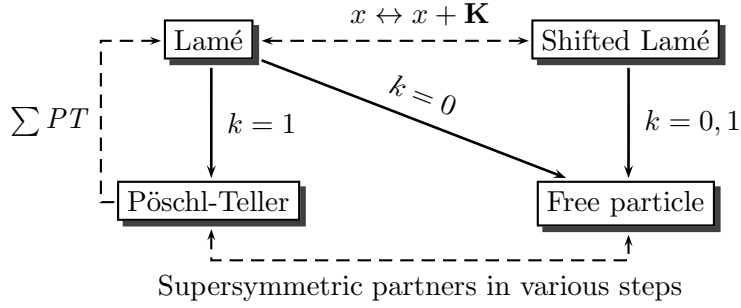

\begin{center}
\begin{equation*}
\begin{psmatrix}
\Rnode{A}{\psshadowbox{\text{Lam\'{e}}}} & & \Rnode{B}{\psshadowbox{\text{Shifted Lam\'{e}}}} \\
\Rnode{C}{\psshadowbox{\text{P\"oschl-Teller}}}& &
\Rnode{D}{\psshadowbox{\text{Free particle}}}
\ncLine[linestyle=dashed,nodesep=1pt]{<->}{A}{B}\naput*{x\leftrightarrow
x+\textbf{K}}
%\ncangle[nodesep=1pt,angleA=90,angleB=-90,armA=0cm,armB=1cm]{<-}{c}{A}\naput*{k=1}
%\ncangle[nodesep=1pt,angleA=90,angleB=-90,armA=0cm,armB=1cm]{<-}{d}{A}
\ncangle[linestyle=dashed,nodesep=1pt,angleA=-90,angleB=-90,armA=0.6cm,armB=0.34cm]{<->}{C}{D}\nbput*{\text{Supersymmetric
partners in various steps}}
\ncangle[linestyle=dashed,nodesep=1pt,angleA=180,angleB=180,armA=3cm,armB=0.8cm]{->}{C}{A}\naput*{\sum
\textit{PT}}
%\ncarc[nodesep=1pt,angleA=90,angleB=-90,armA=0cm,armB=1cm]{<-}{C}{D}
%\ncdiagg[angleA=81,armA=0cm,nodesepA=1pt,nodesepB=1pt,linewidth=1pt]{<-}{f}{B}
%\sum_{l=-\infty}^{\infty }\textit{PT}
 \ncline[nodesep=1pt,linewidth=1pt]{<-}{C}{A}\nbput*{k=1}
\ncline[nodesep=1pt,linewidth=1pt]{->}{A}{D}\naput[nrot=:U]{k=0}
\ncline[nodesep=1pt,linewidth=1pt]{<-}{D}{B}\nbput*{k=0,1}
%\ncline[linestyle=dashed,nodesep=1pt,linewidth=0.7pt]{<-}{C}{D}
%\nccurve[linestyle=dashed,nodesep=1pt,linewidth=1pt,angleA=-135,angleB=135]{<-}{A}{C}
%\ncangle[nodesep=1pt,angleA=90,angleB=-90,armA=0cm,armB=1cm,linewidth=1pt]{<-}{g}{B}\nbput*{k=1}
%\nccurve[linestyle=dashed,angleA=45,angleB=135]{<->}{h}{e}
\end{psmatrix}
\end{equation*}
\end{center}
\bigskip
\caption{The relationship between \textit{PT}, Lam\'e and free
particle
systems.} %\label{related fig}
\end{figure}

\begin{figure}[h]
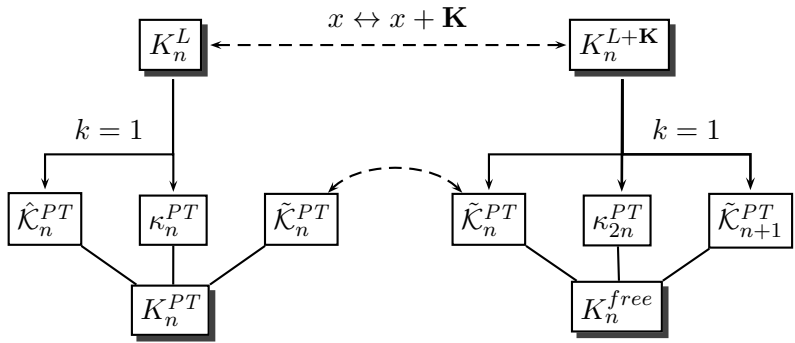

\begin{center}
\begin{equation*}
\begin{psmatrix}
\Rnode{A}{\psshadowbox{K_{n}^{L}}} & \Rnode{B}{\psshadowbox{K_{n}^{L+\mathbf{K}}}} \\
\psset{framesep=0.1}\pstree[levelsep=1.2cm,radius=0.1pt,treemode=U]
{\TR{\psshadowbox{K_{n}^{\textit{PT}}}}}
{\TR{\rnode{c}{\psframebox{\mathcal{\hat{K}}_{n}^{\textit{PT}}}}}
 \TR{\rnode{d}{\psframebox{\mathcal{\kappa }_{n}^{\textit{PT}}}}}
 \TR{\rnode{h}{\psframebox{\mathcal{\tilde{K}}_{n}^{\textit{PT}}}}}} & \psset{framesep=0.1}
\pstree[levelsep=1.2cm,treemode=U,radius=0.1pt]
{\TR{\psshadowbox{K_{n}^{free}}}}
{\TR{\rnode{e}{\psframebox{\mathcal{\tilde{K}}_{n}^{\textit{PT}}}}}
\TR{\rnode{f}{\psframebox{\mathcal{\kappa }_{2n}^{\textit{PT}}}}}
\TR{\rnode{g}{\psframebox{\mathcal{\tilde{K}}_{n+1}^{\textit{PT}}}}}}
\ncLine[linestyle=dashed,nodesep=1pt]{<->}{A}{B}\naput*{x\leftrightarrow
x+\textbf{K}}
\ncangle[nodesep=1pt,angleA=90,angleB=-90,armA=0cm,armB=1cm]{<-}{c}{A}\naput*{k=1}
\ncangle[nodesep=1pt,angleA=90,angleB=-90,armA=0cm,armB=1cm]{<-}{d}{A}
\ncangle[nodesep=1pt,angleA=90,angleB=-90,armA=0cm,armB=1cm]{<-}{e}{B}
%\ncarc[nodesep=1pt,angleA=90,angleB=-90,armA=0cm,armB=1cm]{<-}{f}{B}
\ncdiagg[angleA=81,armA=0cm,nodesepA=1pt,nodesepB=1pt,linewidth=1pt]{<-}{f}{B}
%\ncline[nodesep=1pt,linewidth=1pt]{<-}{f}{B}
\ncangle[nodesep=1pt,angleA=90,angleB=-90,armA=0cm,armB=1cm,linewidth=1pt]{<-}{g}{B}\nbput*{k=1}
\nccurve[linestyle=dashed,angleA=45,angleB=135]{<->}{h}{e}
\end{psmatrix}
\end{equation*}
\end{center}
\caption{Supercharge kernels of $k=1$ Lam\'e, shifted Lam\'e,
\textit{PT} and free particle systems.} \label{kernel fig}
\end{figure}

In the limit $k=1$, Hamiltonian (\ref{isos}) and the supercharge
${Q}_{n}^{L+\mathbf{K}}$ are transformed into
\begin{equation}
H_{n}^{free}=-D^{2}+n^{2}  \label{free}
\end{equation}%
\begin{equation}
Q_{n}^{free}=-iD(D^2-1^2)\ldots (D^2-(n-1)^2)(D^2-n^2).
\label{qfree}
\end{equation}
Operator (\ref{qfree}) is obviously an integral of motion, which is
reduced to a polynomial of order $n$ in Hamiltonian (\ref{free})
multiplied by $D$. Its kernel is spanned by the functions $\cosh
sx$, $\sinh sx$, $s=0,1,\ldots, n$.  In another form the kernel can
be obtained directly from (\ref{lameker}),
\begin{equation}
K_{n}^{L}\underset{x=x+\mathbf{K}}{\longrightarrow }K_{n}^{L+\mathbf{K}}\underset{k=1}{%
\longrightarrow }K_{n}^{free}=\left\{ K_{n-1}^{free},\quad \cosh
^{n}x,~\cosh ^{n-1}\sinh x\right\} ,\quad n>0,\quad K_{0}^{free}=1.
\label{freeker}
\end{equation}
The functions which belong to  (\ref{freeker}) are some linear
combinations of non-normalizable  eigenstates of the Hamiltonian
 (\ref{free}). The special feature of (\ref{freeker}) is that it is composed by functions
 from (\ref{ptker}), in particular, by functions from
 $\mathcal{\tilde{K}}_{n}^{\textit{PT}}$.
Besides, (\ref{freeker}) contains the functions from
$\mathcal{\tilde{K}}_{n+1}^{\textit{PT}}$. Supercharge
(\ref{qfree}) annihilates also  a constant, and the structure of
its kernel (\ref{freeker}) can be summarized as follows,
\begin{equation}
K_{n}^{free}=\left\{ \mathcal{\tilde{K}}_{n}^{\textit{PT}},~\mathcal{\kappa }%
_{2n}^{\textit{PT}},~\left\{ n\,\ \text{functions }\right\} \in ~\mathcal{\tilde{K}}%
_{n+1}^{\textit{PT}}\right\}.
\end{equation}
The relationship between supercharge kernels of $k=1$ Lam\'e,
shifted Lam\'e, \textit{PT} and free particle systems is
presented in Fig. \ref{kernel fig}

As an example, let us consider the simplest case
 $n=1$, for which
the Hamiltonians  (\ref{lame}) and (\ref{isos}) are
\begin{equation}
H_{1}^{L}=-D^{2}+2k^{2}\,\mathrm{sn} ^{2}\,x-k^{2},\qquad
H_{1}^{L+\mathbf{K}}=-D^{2}-2k^{\prime }{}^{2}\,\mathrm{dn}
^{-2}\,x+2-k^{2}. \label{lam1}
\end{equation}
The systems have two allowed bands, and so, three energy
eigenstates associated with band edges. These  states and
corresponding eigenvalues are summarized in the table:
\begin{equation}
\begin{tabular}{|c||c|c|c|}
\hline & Lam\'{e} & Shifted Lam\'{e} & $E_{1,l}$ \\
\hline\hline
$\Psi _{1,0}$ & $\,\mathrm{dn}\,x$ & $1/\mathrm{dn}\,x$ & $0$ \\
\cline{1-2}\cline{3-4}
$\Psi _{1,1}$ & $\,\mathrm{cn}\,x$ & $\mathrm{sn}\,x/\,\mathrm{dn}\,x$ & $%
1-k^{2}$ \\ \cline{1-2}\cline{3-4} $\Psi _{1,2}$ &
$\,\mathrm{sn}\,x$ & $\mathrm{cn}x/\,\mathrm{dn}\,x$ & $1$
\\ \hline
\end{tabular}
\label{tablina}
\end{equation}%
The edge band states are zero modes of the corresponding
supercharges $Q_{1}^{L}$ and $Q_{1}^{L+\mathbf{K}}$, where the
latter is obtained from (\ref{P1}), (\ref{efe})  using the
relation $\mathrm{sn} (x+\mathbf{K})=\mathrm{cn } x/\mathrm{dn
}x$. In the limit $k=1$ the Hamiltonian $H_{1}^{L}$ is transformed
into $H_{1}^{\textit{PT}}=-D^{2}-2\, \mathrm{sech}^{2}x+1$. Its
unique bound state $\psi _{1,0}=\mathrm{sech}x$ originates from
the states $\Psi _{1,0}$ and $\Psi _{1,1}$ of the edges of the
contracting valence band. The singlet state of the continuous
spectrum, $\psi _{1,1}=\tanh x$, originates from the state of the
edge of the conductance band, $\Psi _{1,2}^{L}=\,\mathrm{sn}\,x.$
Together with non-normalizable nonphysical state $\cosh x$, they
form the kernel of $Q_{1}^{\textit{PT}}$. The Hamiltonian of the
shifted Lam\'e system is transformed in this limit into
$H_{1}^{free}=-D^{2}+1,$ and its edge band states (\ref{tablina})
are transformed into non-normalizable states of the system
(\ref{free}) with $n=1$. Only in the case $n=1$ the systems
(\ref{lame}) and  (\ref{isos}) form a pair of supersymmetric
partners \cite{dunsuk}. In correspondence with this, in the limit
$k=1$ the \textit{PT} system is a superpartner of a free particle.
At $k=1$, the supercharge $Q_{1}^{L+\mathbf{K}}$ is transformed
into $Q_{1}^{free}=-i(D-1)D(D+1)=H_{1}^{free}D$, and the physical
states which form the kernel (\ref{freeker}) are
\begin{equation}
K_{1}^{free}=\left\{ \cosh x,~~1,~\sinh x\right\} =\left\{ \mathcal{\tilde{K}%
}_{1}^{\textit{PT}},~\mathcal{\kappa }_{2}^{\textit{PT}},~\sinh
x\in ~\mathcal{\tilde{K}} _{2}^{\textit{PT}}\right\}.
\end{equation}
The two physical states spanning the subspace
$\{\mathcal{\hat{K}}_{1}^{\textit{PT}},~\mathcal{\kappa }%
_{1}^{\textit{PT}}\}$ of the supercharge kernel
$K_{1}^{\textit{PT}}$ originate from the supercharge kernel of the
system (\ref{lame}). The missing non-normalizable nonphysical zero
mode $\cosh x\in \mathcal{ \tilde{K}}_{1}^{\textit{PT}}$ is
provided by the shifted Lam\'{e} system in the limit $k=1$. But
the kernel $K_{1}^{free}$ contains two more zero modes, which are
also nonphysical states and which are related with the system
$H_{2}^{\textit{PT}}$. In particular, $\sinh
x\in\mathcal{\tilde{K}}_{2}^{\textit{PT}}$  and a constant
function from $K_{1}^{free}$ corresponds to $\mathcal{\kappa
}_{2}^{\textit{PT}}$. Since
$\dim\mathcal{\tilde{K}}_{2}^{\textit{PT}}=2,$ there is a missing
state to complete the kernel
$\mathcal{\tilde{K}}_{2}^{\textit{PT}}$. It is provided by $
K_{2}^{free}$, and in this way one  can successively continue.

\vskip 0.2cm

We have analysed Lam\'e equation shifting its argument in the half
of the real period. But it seems to be natural also to look what
happens under shifting the argument for the half of the imaginary
period, $x\rightarrow x+ i\mathbf{K}^{\prime }$, as well as under
the ``diagonal" shift,
\begin{equation} x\rightarrow
x+\mathbf{K}+i\mathbf{K}^{\prime}, \label{rt}
\end{equation}
 with taking
subsequently the limits $k=0$ and $k=1$. The results are
summarized in the following table,
\begin{equation*}
\begin{tabular}{|c|c|c|c|c|}
\hline
& $\overset{}{\underset{}{V_{n}^{L}(x)}}$ & $V_{n}^{L}(x+\mathbf{K})$ & $%
V_{n}^{L}(x+i\mathbf{K}^{\prime })$ & $V_{n}^{L}(x+\mathbf{K}+i
\mathbf{K}^{\prime })$ \\
\hline $k$ & $n(n+1)k^{2}\,\mathrm{sn}^{2}x$ & $-n(n+1)k^{\prime
}{}^{2}\, \mathrm{dn}^{-2}(x,k) $ &
$\overset{}{\underset{}{\frac{n(n+1)}{
\mathrm{sn}^{2}x}}}$ & $n(n+1)\frac{\mathrm{dn}^{2}x}{\mathrm{cn}^{2}x}$ \\
\hline $k=1$ & $-n(n+1)\mathrm{sech}^{2}x$ & $b_{2}$ &
$\overset{}{\underset{}{ \frac{n(n+1)}{\sinh ^{2}x}}}$ & $b_{4}$ \\
\hline
$k=0$ & $b_{1}$ & $b_{3}$ & $\overset{}{\underset{}{\frac{n(n+1)}{\sin ^{2}x}}}$ & $%
\frac{n(n+1)}{\cos ^{2}x}$ \\ \hline
\end{tabular}
\end{equation*}
where $b_i$, $i=1,\ldots,4$, are some constants. These potentials
correspond to P\"oschl-Teller, free particle ($b_{i}$), or
P\"oschl-Teller related systems. Potentials of P\"oschl-Teller
related systems have singularities appearing from the poles of
elliptic functions displaced to the real line. For all the systems
the supercharges of the hidden supersymmetry are obtained from the
supercharge (\ref{LamQ}) of the original Lam\'e system. For the
systems with singular potentials supersymmetry is of a fictitious
nature \cite{ficti}. In such systems, the resulting supercharges
commute with corresponding Hamiltonians, but acting on the
physical states they produce non-normalizable singular states
which violate the boundary conditions, and so, are not physical
states.

\section{\protect\bigskip Conclusion}

We have clarified the nature and the origin of the kernel of the
supercharge of the hidden nonlinear supersymmetry of the
P\"oschl-Teller system by investigating the $k=0$ and $k=1$ limits
of the associated periodic Lam\'e, or appropriately translated
Lam\'e equation. We have showed that in spite of the nonphysical
nature of the non-normalizable states, which constitute a part of
the kernel, they encode essential information on the original
hyperbolic \textit{PT} system, and its singular hyperbolic and
trigonometric modifications, and reflect the intimate relation of
the system to a free particle. In particular, it is interesting to
note that under the Wick rotation, which transforms hyperbolic
\textit{PT} into its singular trigonometric counterpart, and
corresponds to a diagonal translation (\ref{rt}) of Lam\'e system
for a half of the complex period, the nature of physical and
nonphysical states of the \textit{PT} supercharge kernel is
interchanged. This effect is related to the duality in Lam\'e model
discussed by Dunne and Shifman \cite{GerS}.

To conclude, having in mind that the nonlinear superalgebraic
structures, given by Eqs. (\ref{poly}), (\ref{Lapoly}) and
(\ref{polyPT}), have a form  of non-degenerate and degenerate
hyperelliptic curves (\ref{hyper}), it would also be very
interesting to understand the essential differences between the
hidden supersymmetry (as well as its origin) in Lam\'e and in
\textit{PT} systems  within the differential geometric framework of
genus $n$ Riemann surfaces\footnote{In a different, but somewhat
related, context, see also Ref. \cite{GerS}.}.

\bigskip The work has been partially supported  by CONICYT, and by FONDECYT
under grant 1050001.

\end{document}